\documentclass[12pt]{article}
\usepackage{graphicx}
\begin{document}
\begin{titlepage}
\title{Perturbations of the Gravitational Energy in the TEGR:
Quasinormal Modes of the Schwarzschild Black Hole}

\author{J. W. Maluf$\,^{(a)}$, S. C. Ulhoa$\,^{(b)}$, 
F. L. Carneiro$\,^{(c)}$ \\
Instituto de F\'{\i}sica \\
Universidade de Bras\'{\i}lia\\
\bigskip
70.919-970 Bras\'{\i}lia DF, Brazil\\
 K. H. C. Castello Branco$\,^{(d)}$ \\
Instituto de Ci\^encias da Educa\c c\~ao\\
Universidade Federal do Oeste do Par\'{a}\\
68040-070 Santar\'em PA, Brazil}
\date{}
\maketitle
\begin{abstract}
We calculate the gravitational energy spectrum of the perturbations
of a Schwarzschild black hole described by quasinormal modes, in 
the framework of the teleparallel equivalent of general relativity
(TEGR). We obtain a general formula for the gravitational energy 
enclosed by a large surface of constant radius $r$, in the region
$m\,<<\,r\,<<\infty$, where $m$ is the mass of the black hole.
Considering the usual asymptotic expression for the perturbed metric
components, we arrive at finite values for the energy spectrum. The
perturbed energy depends on the two integers $n$ and $l$ that describe
the quasinormal modes. In this sense, the energy perturbations are
discretised.  We also obtain a simple expression for the
decrease of the flux of gravitational radiation of the perturbations.
\end{abstract}
\thispagestyle{empty}
\vfill
\noindent (a) jwmaluf@gmail.com, wadih@unb.br\par
\noindent (b) sc.ulhoa@gmail.com\par
\noindent (c) fernandolessa45@gmail.com\par
\noindent (d) khccb@yahoo.com.br\par
\end{titlepage}
\newpage

\section{Introduction}
The response of a black hole or neutron star to external, nonradial 
perturbations is described by quasinormal modes. A comprehensive
review of the physics related to this interesting phenomena 
is found in refs. \cite{Nollert,Kokkotas}. These modes are 
damped oscillations of the space-time geometry that may be used to 
characterize the intrinsic properties of the physical system.
Investigations on quasinormal modes are carried out both analytically
and numerically. The modes are characterized by a spectrum of 
discrete, complex valued frequencies. The real part of the frequency 
is related to the oscillation frequency, and the imaginary part yields
the rate at which each mode is damped as a result of emission of 
radiation. Thus, quasinormal modes may be important to 
gravitational waves astrophysics. In addition, the study of these modes
is, to some extent, a testing ground for ideas in quantum gravity.
Chandrasekhar once stated \cite{Chandra} that one relevant way of 
investigating a physical system is by perturbing it, and analysing the 
response of the system. This is precisely the role of QNM in the
context of black holes.

In the analysis of Einstein's equations for a perturbed Schwarzschild
space-time one finds a number of equations that are eventually reduced
to two one-dimensional wave equations for the perturbed metric
components (for axial and polar perturbations) \cite{Nollert}.
However, the nature of the potential in these equations precludes
exact solutions in terms of known functions \cite{SW}. In particular, 
the equations for the metric perturbations admit solutions
provided the frequencies are discrete and under the imposition of
special boundary conditions. The solutions of these equations are
the quasinormal modes.

In this article, we calculate the energy spectrum of the quasinormal 
modes in the framework of the teleparallel equivalent of general 
relativity (TEGR). A proper definition for the energy-momentum of the
gravitational field is of ultimate importance for a comprehensive
understanding of Einstein's general relativity, and yet there is no 
general agreement regarding an acceptable expression. There is not an
unique approach to the problem (pseudotensors, quasilocal expressions, 
for instance), and even within each approach there is no 
preferred definition for the energy-momentum. In the context of the
field equations of the TEGR, however, one finds a suitable and consistent
framework for the definitions of the gravitational energy-momentum and
4-angular momentum \cite{Maluf1,Maluf4}, that satisfy the algebra of the
Poincar\'{e} group. This definition of the gravitational energy-momentum
has been applied to several configurations of the 
gravitational field, and leads to consistent results. It 
satisfies important requirements that any gravitational 
energy-momentum definition must satisfy \cite{Maluf4}.

We will consider the perturbed Schwarzschild space-time and calculate 
the gravitational energy enclosed by a surface of constant radius $r$.
The energy contained within this region of the unperturbed
space-time is known and may be easily evaluated out of the definition
that arises in the TEGR, as well as by means of several quasilocal
definitions for the gravitational energy. Thus, by subtracting the 
unperturbed energy from the total energy that includes the perturbations,
we obtain the energy of the perturbations only. In the present analysis,
we will consider both axial and polar perturbations. We are restricting
the analysis to gravitational perturbations (we do not address scalar
and electromagnetic perturbations). 
We find that the perturbed energy (i) oscillates and at
the same time is damped, and (ii) depends on the integers $n$ (the
overtone number) and $l$ (the angular momentum number). To our 
knowledge, a similar analysis regarding the spectrum of the energy 
perturbations of the Schwarzschild space-time has not been presented
so far. We conjecture that the perturbed energy in arbitrary 
(nonspherical) volumes of the Schwarzschild space-time is also 
characterized by integers. 

Notation: space-time indices $\mu, \nu, ...$ and SO(3,1)
indices $a, b, ...$ run from 0 to 3. Time and space indices are
indicated according to
$\mu=0,i,\;\;a=(0),(i)$. The tetrad field is denoted $e^a\,_\mu$,
and the torsion tensor reads
$T_{a\mu\nu}=\partial_\mu e_{a\nu}-\partial_\nu e_{a\mu}$.
The flat, Minkowski space-time metric tensor raises and lowers
tetrad indices and is fixed by
$\eta_{ab}=e_{a\mu} e_{b\nu}g^{\mu\nu}= (-1,+1,+1,+1)$. The 
determinant of the tetrad field is represented by 
$e=\det(e^a\,_\mu)$.\par        
\bigskip

\section{Review of the gravitational energy-momentum definition in the
TEGR}

We assume that the space-time geometry is defined by the
tetrad field $e^a\,_\mu$ only. In this case, the only
possible nontrivial definition for the torsion tensor is given by
$T_{a\mu\nu}=\partial_\mu e_{a\nu}-\partial_\nu e_{a\mu}$. In the TEGR, it is
possible to rewrite Einstein's equations in terms of $e^a\,_\mu$ and 
$T_{a\mu\nu}$. The Lagrangian density of the theory is defined by

\begin{eqnarray}
L&=& -k e({1\over 4}T^{abc}T_{abc}+{1\over 2}T^{abc}T_{bac}-
T^aT_a) -{1\over c}L_M \nonumber \\
&\equiv& -ke\Sigma^{abc}T_{abc} -{1\over c}L_M\,, 
\label{1}
\end{eqnarray}
where $k=c^3/16\pi G$, $T_a=T^b\,_{ba}$, 
$T_{abc}=e_b\,^\mu e_c\,^\nu T_{a\mu\nu}$ and

\begin{equation}
\Sigma^{abc}={1\over 4} (T^{abc}+T^{bac}-T^{cab})
+{1\over 2}( \eta^{ac}T^b-\eta^{ab}T^c)\;.
\label{2}
\end{equation}
$L_M$ stands for the Lagrangian density for the matter fields. 
The Lagrangian density $L$ is invariant under the global SO(3,1)
group. Invariance under the local SO(3,1) group is verified as long as
we take into account the total divergence that arises in the identity 

$$
eR(e) \equiv -e\left({1\over 4}T^{abc}T_{abc} + 
{1\over 2}T^{abc}T_{bac} - T^{a}T_{a}\right)
+ 2\partial_{\mu}(eT^{\mu})\,.
$$
where $R(e)$ is the scalar Riemannian curvature. However, the field 
equations derived from eq. (\ref{1}) are invariant under local
SO(3,1) transformations, and are equivalent to Einstein's equations. 
They read

\begin{equation}
e_{a\lambda}e_{b\mu}\partial_\nu (e\Sigma^{b\lambda \nu} )-
e (\Sigma^{b\nu}\,_aT_{b\nu\mu}-
{1\over 4}e_{a\mu}T_{bcd}\Sigma^{bcd} )={1\over {4kc}}e\texttt{T}_{a\mu}\,,
\label{3}
\end{equation}
where
$\delta L_M / \delta e^{a\mu}=e\texttt{T}_{a\mu}$. 

The definition of the gravitational energy-momentum may be 
established in the framework of the Lagrangian formulation defined by
(\ref{1}), according to the procedure of ref. \cite{Maluf1} (we will make
$c=1=G$). Equation (\ref{3}) may be rewritten as 

\begin{equation}
\partial_\nu(e\Sigma^{a\lambda\nu})={1\over {4k}}
e\, e^a\,_\mu( t^{\lambda \mu} + \texttt{T}^{\lambda \mu})\;,
\label{4}
\end{equation}
where $\texttt{T}^{\lambda\mu}=e_a\,^{\lambda}\texttt{T}^{a\mu}$ and
$t^{\lambda\mu}$ is defined by

\begin{equation}
t^{\lambda \mu}=k(4\Sigma^{bc\lambda}T_{bc}\,^\mu-
g^{\lambda \mu}\Sigma^{bcd}T_{bcd})\,.
\label{5}
\end{equation}
In view of the antisymmetry property 
$\Sigma^{a\mu\nu}=-\Sigma^{a\nu\mu}$, it follows that

\begin{equation}
\partial_\lambda
\left[e\, e^a\,_\mu( t^{\lambda \mu} + \texttt{T}^{\lambda \mu})\right]=0\,.
\label{6}
\end{equation}
The equation above yields the continuity (or balance) equation,

\begin{equation}
{d\over {dt}} \int_V d^3x\,e\,e^a\,_\mu (t^{0\mu} +\texttt{T}^{0\mu})
=-\oint_S dS_j\,
\left[e\,e^a\,_\mu (t^{j\mu} +\texttt{T}^{j\mu})\right]\,.
\label{7}
\end{equation}
Therefore we identify
$t^{\lambda\mu}$ as the gravitational energy-momentum tensor
\cite{Maluf1},

\begin{equation}
P^a=\int_V d^3x\,e\,e^a\,_\mu (t^{0\mu} 
+\texttt{T}^{0\mu})\,,
\label{8}
\end{equation}
as the total energy-momentum contained within a volume $V$,

\begin{equation}
\Phi^a_g=\oint_S dS_j\,
\, (e\,e^a\,_\mu t^{j\mu})\,,
\label{9}
\end{equation}
as the gravitational energy-momentum flux \cite{Maluf2,Maluf3}, and

\begin{equation}
\Phi^a_m=\oint_S dS_j\,
\,( e\,e^a\,_\mu \texttt{T}^{j\mu})\,,
\label{10}
\end{equation}
as the energy-momentum flux of matter \cite{Maluf3}. In view of (\ref{4}),
eq. (\ref{8}) may be written as 

\begin{equation}
P^a=-\int_V d^3x \partial_j \Pi^{aj}=-\oint_S dS_j\,\Pi^{aj}\,,
\label{11}
\end{equation}
where $\Pi^{aj}=-4ke\,\Sigma^{a0j}$. A summary of all issues discussed above
may found in ref. \cite{Maluf9}

Equation (\ref{11}) is the 
definition for the gravitational energy-momentum presented in ref.
\cite{Maluf4}, obtained in the framework of the vacuum field
equations in Hamiltonian form. It is invariant under coordinate 
transformations of the three-dimensional space and under time 
reparametrizations. Note that (\ref{6}) is a true energy-momentum conservation
equation. We also remark that for finite volumes of integration, the free 
index of the integral of the energy-momentum 4-vector density must be a 
Lorentz index, such as $a$ in the left hand side of definition (\ref{11}), 
because under a global SO(3,1) transformation both sides of (\ref{11}) 
transform consistently. In contrast, the integral on the right hand side of a
hypothetical quantity such as $P^\mu =\int d^3x\,e V^\mu$ 
(with space-time index $\mu$) would be coordinate dependent (the integral of
a space-time vector density $eV^\mu$ is neither 
invariant nor covariant under coordinate transformations), and therefore 
ill-defined for finite, three-dimensional volumes of integration.

In the ordinary formulation of arbitrary field theories, energy, 
momentum, angular momentum and the centre of mass
moment are frame dependent field quantities, that 
transform under the global SO(3,1) group. In particular, energy 
transforms as the zero component of the energy-momentum four-vector. 
These features of special relativity must also hold in general relativity,
since the latter yields the former in the limit of weak (or vanishing) 
gravitational fields. As 
an example, consider the total energy of a black hole, represented by the 
mass parameter $m$. As seen by a distant observer, the total energy of a 
static Schwarzschild black hole is given by $E = mc^2$. However, at
great distances the black hole may be considered as a particle of mass $m$,
and if it moves with constant velocity $v$, then its total energy as seen by
the same distant observer is $E = \gamma mc^2$, where 
$\gamma=(1-v^2/c^2)^{-1/2}$. Likewise, the gravitational momentum, angular 
momentum and the centre of mass moment are naturally frame dependent field 
quantities, whose values vary from frame to frame, and are different for
different observers in arbitrary space-times.

Before closing this section, we note that the existence of a scalar density
(on a three-dimensional spacelike hypersurface)
like $\partial_j\Pi^{aj}=-4k\partial_j(e\,\Sigma^{a0j})$ in eq. (\ref{11}) 
is natural in the framework of theories constructed out of third order tensors 
such as the torsion tensor $T_{a\mu\nu}$. Naively, we observe that the 
contraction of two space-time indices in a third order
tensor yields a vector of the type $\phi^\mu$, and therefore
$\partial_\mu (e\phi^\mu)$ is a well defined scalar density that under 
integration yields a well behaved surface integral, in similarity to
eq. (\ref{11}). The existence of these well defined scalar densities is 
natural in the TEGR, as in the identity below eq. (2), but of course, after a
number of manipulations and rearrangements, these scalar densities could also
be established in Einstein-Cartan type theories.

\section{ Axial perturbations of the Schwarzschild black hole}

The analysis of the quasinormal modes in the Schwarzschild space-time
consists in solving the differential equations for the perturbed
metric components, with appropriate boundary conditions.
These equations were first written down by Regge 
and Wheeler \cite{RW}. They obtained the general form of the simplest
nonspherical perturbations for the Schwarzschild black hole, namely, 
the axial and polar
perturbations. Regge and Wheeler showed that the equations that 
describe the axial perturbations may be separated if the perturbed 
metric tensor $h_{\mu\nu}$ is expanded in tensorial spherical 
harmonics. The general form of $h_{\mu\nu}$ is obtained and
further simplified by taking into account the gauge symmetry (coordinate
invariance) of the field equations. The simplest form of the 
nonvanishing axial perturbations are given by \cite{RW}

\begin{equation}
h_{03}=h_0(t,r)\sin\theta\,\partial_\theta P_l(\theta)\,,
\hskip 1.0cm
h_{13}=h_1(t,r)\sin\theta\,\partial_\theta P_l(\theta)\,.
\label{12}
\end{equation}
where $P_l(\theta)$ are the Legendre polynomials. 
The functions $h_0(t,r)$ and $h_1(t,r)$ satisfy the 
equations \cite{Nollert,RW}

\begin{eqnarray}
{1\over {f(r)}}{{\partial h_0}\over {\partial t}}-
{{\partial \lbrack f(r)h_1\rbrack}\over {\partial r}}&=&0\,, 
\label{13} \\
{1\over {f(r)}}\biggl[ {{\partial^2h_1}\over {\partial t^2}}-
{{\partial^2h_0}\over {\partial t \partial r}} 
+{2\over r} {{\partial h_0}\over {\partial t}}
\biggr]
+ {1\over r^2}\lbrack l(l+1)-2\rbrack h_1&=&0 \,, 
\label{14} \\
{f(r) \over 2}\biggl[
{{\partial^2 h_0}\over {\partial r^2}}-
{{\partial^2 h_1}\over {\partial t \partial r}} -
{2\over r} {{\partial h_1}\over {\partial t}} \biggr]+
{1\over r^2} \biggl[ r{{\partial f}\over {\partial r}}-
{1\over 2}l(l+1)\biggr]h_0&=&0\,,\label{15}
\end{eqnarray}
where $f(r)=1-2m/r$. Equation (\ref{15}) is a consequence of (\ref{13}) and
(\ref{14}). The quantity $\partial h_0/\partial t$ in (\ref{13}) may be
substituted in (\ref{14}). Thus, (\ref{13}) and (\ref{14}) may be combined into
one equation. Defining $\Psi(t,r)=(1/r) f(r) h_1(t,r)$, the resulting equation
reads

\begin{equation}
{{\partial^2 \Psi}\over{\partial t^2}}-
{f\over r}{\partial \over {\partial r}}\biggl[
f {{\partial (r\Psi)}\over {\partial r}}\biggr]+
{{2f^2}\over r^2} {{\partial (r\Psi)}\over {\partial r}}+
{f\over r^2}\lbrack l(l+1)-2 \rbrack \Psi=0\,.
\label{16}
\end{equation}
In terms of the tortoise coordinate 
$x=r+2m\ln ( r/2m-1 )$, eq. (\ref{16}) may be rewriten as

\begin{equation}
{{\partial \Psi^2}\over {\partial t^2}}-
{{\partial\Psi^2}\over{\partial x^2}}+V_{RW}(x)\Psi(t,x)=0\,.
\label{17}
\end{equation}
$V_{RW}(x)$ is the Regge-Wheeler potential. It is given by

\begin{equation}
V_{RW}(x)=\biggl(1-{{2m}\over r}\biggr)
\biggl( {{l(l+1)}\over {r^2}}- {{6m}\over r^3}\biggr)\,,
\label{18}
\end{equation}
and $x=x(r)$.

The time dependence of $h_{\mu\nu}$ is assumed to be of the type
$e^{-i\omega t}$, where $\omega$ is a complex frequency. Thus, it follows
that $\Psi(t,r)=e^{-i\omega t} \psi(r)$. As a consequence, the 
function $\psi(r)$ satisfies the time independent equation

\begin{equation}
{{d^2 \psi}\over {dr^2}}+\lbrack \omega^2 -V_{RW}(x)\rbrack \psi(x)=0\,.
\label{19}
\end{equation}
In view of the asymptotic behaviour of the potential $V_{RW}(x)$, the 
expressions of $\psi(x)$ in the asymptotic limits 
$x\rightarrow \infty\;\; (r\rightarrow \infty)$ and
$x\rightarrow-\infty\;\; (r \rightarrow 2m)$ are given by
$\psi(x)\cong Ae^{ i\omega x}$ and $\psi(x)\cong Ae^{- i\omega x}$,
respectively, where $A<<1$ in order to ensure the perturbative
character of the solution. Expressions with positive exponential correspond
to waves that are purely outgoing at infinity, and the ones with
negative exponential, to waves that are purely ingoing at the horizon.

For weakly excited states, the quasi-normal frequencies $\omega$ may be 
obtained
from the third order WKB approximation. In the case of the gravitational
perturbation, the fundamental state is given by $n=0$ and $l=2$ 
\cite{Iyer-Will,iyer1987black}. These semi-analytic approximations yield 
values very close
to the numerical values \cite{konoplya2003quasinormal}. The stable solutions
decay with time. Thus, if $\omega_r$ and $\omega_i$ represent the real and
imaginary parts, respectively, of the complex valued frequency, then we must
have $\omega_{i}<0$. 

The possible values of $\omega$ for highly damped modes (large values
of $n$) are independent of $l$, and may be obtained by means of 
numerical investigations. They are given by \cite{Nollert}

\begin{equation}
\omega_n= {{\ln 3}\over {8 \pi m}} - {i\over {4m}}
\biggl(n+{1\over 2}\biggr) +O(n^{-1/2})\,,
\label{20}
\end{equation}
where $n$ is an integer. This expression has also been obtained by means of
analytic procedure in ref. \cite{Berti}. However, the frequency to be 
employed in the construction of the Figures below, in subsections 3.1 and 4.1,
does not correspond to highly damped modes.

The problems regarding the
divergent behaviour of $h_1(t,r)$ in the limit 
$r\rightarrow \infty$ have been discussed in refs. 
\cite{Nollert} (section 3.1.2) and \cite{Nollert2}. 
However, the asymptotic behaviour of 
$h_1(t,r)$ poses no problem to the result of our analysis, since we
will integrate (\ref{11}) over a surface of constant (finite) radius 
$r<<\infty$.

We need the asymptotic behaviour of $h_0(t,r)$. Given that
$\partial_0 h_0=i\omega\,h_0$, and

\begin{equation}
\partial_0 h_0
=f{{\partial \lbrack f(r)h_1\rbrack}\over {\partial r}}
\cong \partial_1 h_1 \,,
\label{21}
\end{equation}
which follows from (\ref{13}) in the limit $r\rightarrow \infty$, we find

\begin{equation}
h_{0}\approx\left(\frac{if}{\omega}-r\right)\Psi\approx-r\Psi\,.
\label{22}
\end{equation}
The $m/r$ correction to the expressions above will be negligible in
the development of our analysis. By means of straightforward 
calculations, we obtain from  (\ref{22}) an expression that
holds in the limit $r\rightarrow \infty$ and will be useful in the 
following subsection. We find 

\begin{equation}
\left(\partial_{1}h_{0}-\partial_{0}h_{1}\right)\approx 
-\frac{2}{r}h_{1}\approx-2\Psi\,.
\label{23}
\end{equation}

\subsection{The gravitational energy of the axial perturbations}

In this subsection, we will address the axially perturbed Schwarzschild 
space-time and calculate the gravitational energy contained within a simple
three dimensional volume, namely, a spherical volume of radius $r$.
In order to simplify the calculations, we will consider a large
sphere so that the field quantities are considered in the limit
$m/r <<1$. The purpose is to find the expression of the
energy perturbations in this limit. To our knowledge, no expression 
for the gravitational energy perturbations has been obtained so far
in the present physical context.

In the course of the calculations we find that the perturbed
gravitational energy depends on the square of the field quantities 
$h_0$ and $h_1$ that appear in (\ref{12}). Therefore in order to calculate
the first order contribution to the perturbed energy we must keep
the terms quadratic in $h_{03}$ and $h_{13}$. 

We start with the metric tensor \cite{RW}

\begin{eqnarray}
ds^2&=& -\biggl(1-{{2m}\over r}\biggr)dt^2+
\biggl(1-{{2m}\over r}\biggr)^{-1}dr^2+ +r^2d\theta^2+
r^2\sin^2\theta d\phi^2 \nonumber \\
&{}&+2h_0\sin\theta \,\partial_\theta P_l(\cos\theta)\,dt d\phi+
2 h_1\sin\theta \,\partial_\theta P_l(\cos\theta)\, dr d\phi\,.
\label{24}
\end{eqnarray}
Definition (\ref{11}) for the gravitational energy-momentum is frame 
dependent. Thus, we choose a configuration of tetrad fields that has 
a clear physical interpretation. Tetrad fields are interpreted as 
reference frames adapted to preferred fields of observers in
space-time. This interpretation is possible by identifying the 
$e_{(0)}\,^\mu$ components of the frame with the four-velocities
$u^\mu$ of the observers, $e_{(0)}\,^\mu=u^\mu$ 
\cite{Maluf5,Hehl2}. In the present 
analysis we will establish a set of tetrad fields adapted to static
observers in space-time. Therefore, we require $e_{(0)}\,^i=0$. This
condition fixes 3 components of the frame. The other three
components are fixed by choosing a orientation of the frame in the
three-dimensional space. Thus, $e_{(0)}\,^\mu$ is parallel to
the worldline of the observers, and  $e_{(k)}\,^\mu$ are the three
unit vectors orthogonal to the timelike direction. We fix 
$e_{(k)}\,^\mu$ such that $e_{(1)}\,^\mu$, $e_{(2)}\,^\mu$ and 
$e_{(3)}\,^\mu$ in cartesian coordinates are unit vectors along 
the $x$, $y$ and $z$ directions. The tetrad field in 
$(t,r,\theta,\phi)$ coordinates that satisfies these conditions is 
given by

\begin{equation}
e_{a\mu}=\pmatrix{
-A&0&0&-D\cr
0&B\sin\theta\cos\phi & r\cos\theta\cos\phi &
-Er\sin\theta\sin\phi+F\sin\theta\cos\phi\cr
0&B\sin\theta\sin\phi & r\cos\theta\sin\phi &
 Er\sin\theta\cos\phi+F\sin\theta\sin\phi\cr
0&B\cos\theta & -r\sin\theta & F \cos\theta}\,,
\label{25}
\end{equation}
with the following definitions:

\begin{eqnarray}
A&=& (-g_{00})^{1/2}\,, \nonumber \\
B&=& (g_{11})^{1/2} \,, \nonumber \\
D&=& -{{g_{03}}\over {(-g_{00})^{1/2}}}\,, \nonumber \\
E^2&=& 1+
{{g_{11}g_{03}^2+g_{00}g_{13}^2}\over {r^2\sin^2\theta}}\,, 
\nonumber \\
F&=& {{g_{13}}\over {(g_{11})^{1/2}}}\,.
\label{26}
\end{eqnarray}
The frame above satisfies $e_{(0)}\,^i=0$.
It is possible to show that if we neglect $g_{03}$ and $g_{13}$, the 
frame components $e_{(k)}\,^\mu(t,x,y,z)$ in the limit $m/r<<1$
are given by $e_{(1)}\,^\mu \cong (0,1,0,0)\,$,
$e_{(2)}\,^\mu \cong (0,0,1,0)\,,$ and 
$e_{(3)}\,^\mu \cong (0,0,0,1)$.

We denote by $g$ and $e$ the determinants of $g_{\mu\nu}$ and
$e^a\,_{\mu}$, respectively. We find $g=-e^2=-r^4\sin^2\theta\,E^2$.
Thus, we have
\begin{equation}
e=r^{2}\sin{\theta}\sqrt{1+f^{-1}
\frac{g_{03}^2}{r^{2}\sin^{2}{\theta}}-
f\frac{g_{13}^2}{r^{2}\sin^{2}{\theta}}}\,.
\label{27}
\end{equation}

The gravitational energy is given by the $a=(0)$ component of (\ref{11}).
Transforming the volume integral into a surface integral and 
considering the definition $\Pi^{aj}=-4k\,\Sigma^{a0j}$, we find that

\begin{equation}
P^{(0)}=-\oint_S dS_j\,\Pi^{(0)j}=\oint_SdS_j\,4ke\,\Sigma^{(0)0j}\,.
\label{28}
\end{equation}
The integration will be carried out on a surface $S$ of constant 
radius $r$. 

The non-vanishing components of the torsion tensor that are relevant
to the evaluation of the gravitational energy are

\begin{eqnarray}
T_{001}&=& A\partial_1 A \,,\nonumber \\
T_{301}&=& D\partial_1 A \,,\nonumber \\
T_{003}&=& -A\partial_0 D \,,\nonumber \\
T_{103}&=&  B\partial_0 F \,,\nonumber \\
T_{303}&=& rE\partial_0 (rE)\,\sin^2\theta-D\partial_0 D
+F\partial_0 F\,,\nonumber \\
T_{013}&=& -A\partial_1 D\,, \nonumber \\
T_{113}&=& B\partial_1 F\,, \nonumber \\
T_{212}&=& r(1-B)\,, \nonumber \\
T_{313}&=& rE(\partial_1(rE)-B)\sin^2\theta-D\partial_1D
+F\partial_1 F \,,\nonumber \\
T_{223}&=& rF\,.
\label{29}
\end{eqnarray}
The evaluation of $\Sigma^{(0)01}$ requires a large
number of algebraic manipulations, but otherwise is simple. The full
expression of $e\Sigma^{(0)01}$ is given by

\begin{eqnarray}
e\Sigma^{(0)01}&=&
-r\sin\theta\sqrt{
1+{1\over f} {{g_{03}^2}\over {r^2\sin^2 \theta}}
-f{{g_{13}^2}\over {r^2\sin^2} \theta}}\times \nonumber \\
&{}&
\biggl\{2(f^{1/2}-1)g_{03}^2+2f\biggl[ fg_{13}^2+2r^2f^{1/2}\sin^2\theta
\nonumber \\
&{}&-r^2\sin^2\theta\biggl(
1+\sqrt{
1+{1\over f} {{g_{03}^2}\over {r^2\sin^2 \theta}}
-f{{g_{13}^2}\over {r^2\sin^2} \theta}}\biggr)\biggr]\nonumber \\
&{}& +rf^{1/2}g_{03}(\partial_1 g_{03}-\partial_0 g_{13})\biggr\}\times
\nonumber \\
&{}&\biggl\{ 4\bigl[g_{03}^2+f\bigl(-fg_{13}^2+r^2\sin^2\theta\bigr)\bigr]
\biggr\}^{-1}\,.
\label{30}
\end{eqnarray}

By keeping terms of order $(h_0)^2$ and $(h_1)^2$ 
only in the integrand of (\ref{30}),
we obtain the approximate expression that is relevant to the present analysis.
The leading term in the energy expression is $m$, the
total Schwarzschild energy. It is easy to verify that 

\begin{equation}
4k \int_{r\rightarrow \infty} d\theta \,d\phi \,
\lbrack -\sin\theta(-g_{00})^{1/2}\,r(1-B)\rbrack =m\,,
\label{31}
\end{equation}
which is a well known result. We recall that $B$ is given by (\ref{26}).

The gravitational energy of the perturbations will be denoted by 
$\delta P^{(0)}$, and is obtained out of the second order terms in (\ref{30}),
assuming $-g_{00}\cong 1$ and $g_{11}\cong 1$.
Therefore, for fixed values of $r$ such that $m/r <<1$ we obtain,
after some simplifications,

\begin{equation}
\delta P^{(0)} =k \int d\theta d\phi\,
\frac{\sin{\theta}}{r}(P_{l}(\cos{\theta}))^{2}
\left[h_{0}^{2}-3h_{1}^{2}-rh_{0}
\left(\partial_{1}h_{0}-\partial_{0}h_{1}\right)\right]\,.
\label{32}
\end{equation}
Recalling that $k=1/16\pi$ (we are assuming $c=1=G$), we find

\begin{equation}
\delta P^{(0)}={1\over 8}\int d\theta \,
\frac{\sin{\theta}}{r}(P_{l}(\cos{\theta}))^{2}
\left[h_{0}^{2}-3h_{1}^{2}-rh_{0}
\left(\partial_{1}h_{0}-\partial_{0}h_{1}\right)\right]\,.
\label{33}
\end{equation}

\begin{figure}[htbp]
	\centering
		\includegraphics[width=1\textwidth]{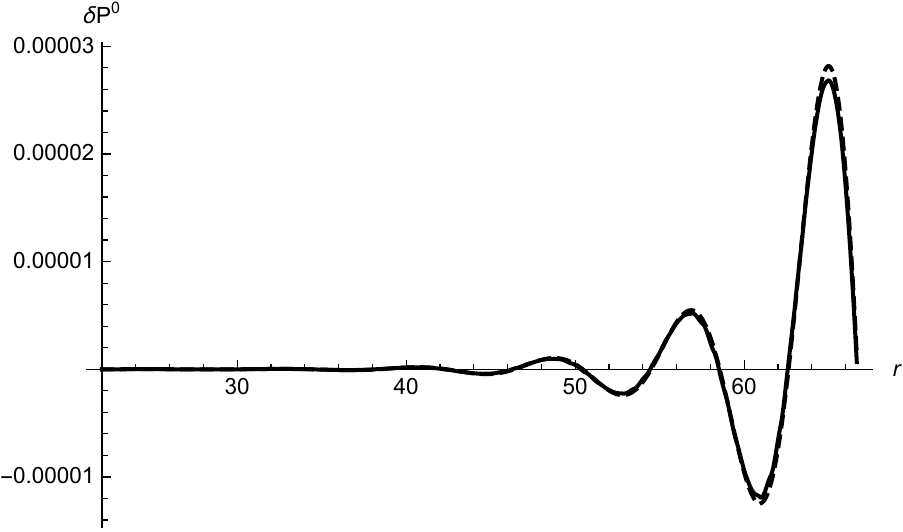}
	\caption{Comparison between the real parts of the gravitational energy of
	the axial perturbation. The continuous line represents the values of the 
	numerical integration of the 
	energy, whereas the dashed line represents the values obtained from the
	approximate analytic expression (\ref{34}). The parameters used are
	$A=10^{-6}$, $m=1$. The frequency was required to be the fundamental
	frequency ($n=0,l=2$)$\;$ $\omega=0.373162-i0.089217$, obtained from the third
	order WKB method. The data represent the initial instant of time of the
	perturbation, i.e., $t=0$.}\label{fig1}
\end{figure}

Now we make use of (\ref{22}) and (\ref{23}), and we arrive at
\begin{equation}
\delta P^{(0)}\approx -\frac{l(l+1)}{2l+1}r\Psi^{2}\label{34}\,,
\end{equation}
where we have used 

$$\int_{0}^{\pi} d\theta\sin{\theta}\partial_{\theta}(P_{l}(\cos{\theta}))^{2}
=\frac{2l(l+1)}{2l+1}\,.$$
In the absence of the perturbation, i.e., when $h_0=0=h_1$, 
$\delta P^{(0)}$ vanishes.
The approximate solution (\ref{34}) may be compared with the numerical solution
presented in Figure \ref{fig1}. The numerical solution, represented by the 
continuous line, is obtained directly from the exact expression for the total
gravitational energy minus the black hole
energy, requiring the function $\psi$ to be a solution of eq. 
(\ref{19}). The boundary conditions for the numerical 
solution of eq. (\ref{19}) are $\psi(r_1)=A \exp^{-i \omega x_1}$ and 
$\psi(r_2)=A \exp^{+i \omega x_2}$, where 
$x_1=r_1+2 m \ln(r_1/2m-1)$, $x_2=r_2+2 m \ln(r_2/2m-1)$,  
$r_{1}=2m+m/4$ and $r_{2}=200m$.

\begin{figure}[htbp]
	\centering
		\includegraphics[width=1\textwidth]{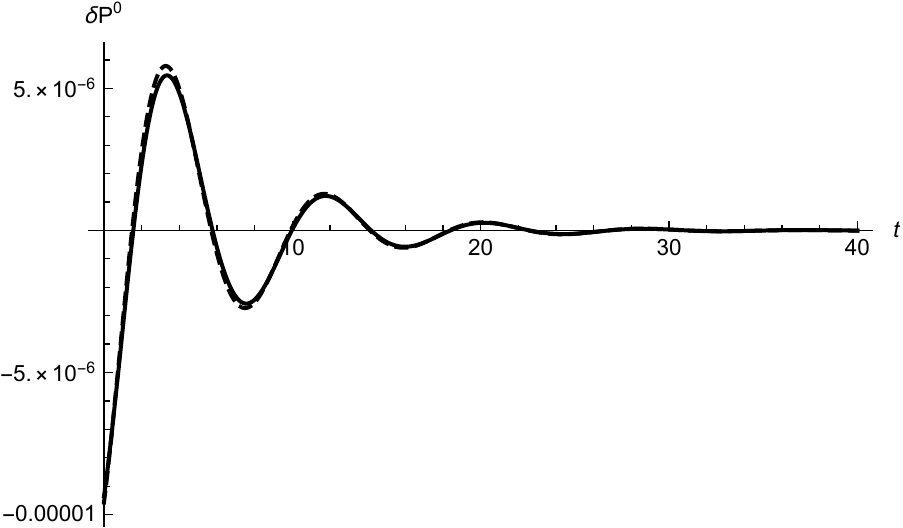}
	\caption{Comparison between the real parts of the gravitational energy of the
	axial perturbation, as a function of time. The continuous line 
	represents the values of the numerical	integration of the 
	energy, whereas the dashed line represents the values obtained from the
	approximate analytic expression (\ref{34}). The parameters used are
	$A=10^{-6}$, $m=1$. The frequency was required to be the fundamental
	frequency ($n=0,l=2$)$\;$ $\omega=0.373162-i0.089217$, obtained from the third
	order WKB method. The radius of integration was chose to be 
	$r=60$, in natural units.}\label{fig2}
\end{figure}

In Figure \ref{fig1} we display the initial instant of time of the 
perturbation. We see that the larger is the value of the coordinate $r$,
more intense is the energy of the perturbation, probably because the farther
one is from the black hole, the more energy is necessary to perturb it.
Note that $r$ is not a dynamical
coordinate, it is just a parameter of the coordinate system that fixes the 
radius of integration. The evolution of the gravitational energy perturbation
is stable, i.e., the latter tends to zero when $t\rightarrow \infty$, as we see
in Figure \ref{fig2}.


Considering the real part of the solution given by eq. (\ref{34}), we have

\begin{equation}\label{35}
\delta P^{(0)}=-\frac{l(l+1)}{2l+1}rA^{2}
\cos{\left[2\omega_{r}(x-t)\right]}e^{-2\omega_{i}(x-t)}\,.
\end{equation}

We may calculate the variation of the perturbation energy in a half-life 
interval $T/2$, for weakly damped solutions, but neglecting the damping effects,
considering only $\omega=2\omega_{r}$. By making use of 

\begin{equation}\label{36}
\frac{1}{T/2}\int^{T/2}_{0}dt\sin\left(\omega t\right)=\frac{2}{\pi}\,,
\end{equation}
we obtain
\begin{equation}\label{37}
\delta P^{(0)}_{1/2}=-\frac{2rA^{2}}{\pi}\frac{l(l+1)}{2l+1}\,.
\end{equation}

Since the black hole looses energy in the course of time, as we see in 
Figure (\ref{fig2}), it is important
to analyse the rate at which the energy decay occurs. This issue may be
described by the energy flux of the radiated energy.
The present formalism for the gravitational energy-momentum allows
a definition for the gravitational energy-momentum flux \cite{Maluf3},
which has been applied to the radiation of gravitational
energy in the Bondi space-time.
The gravitational energy flux is given by the $a=(0)$ component of
$\Phi^a_g$ defined by (\ref{9}), or simply by

\begin{equation}
\Phi^{(0)}_g=-{{dP^{(0)}}\over {dt}}
=-{{d(\delta P^{(0)})}\over {dt}}\,.
\label{44}
\end{equation}
In the present context, we find

\begin{eqnarray}
\Phi^{(0)}_g&=&
\frac{2l(l+1)}{2l+1}A^2 r\biggl[
\omega_r \sin\lbrack 2\omega_r(x-t)\rbrack \nonumber \\
&{}&+\omega_i \cos\lbrack 2\omega_r(x-t) \rbrack 
\biggr]\,e^{-2\omega_i(x-t)}\,.
\label{39}
\end{eqnarray}
Again we define $\omega=2\omega_r$ and find that after a period
$T=2\pi/\omega$ the decrease in the flux of gravitational radiation
is $\Phi^{(0)}_g(t+T)-\Phi^{(0)}_g(t)=
-\Phi^{(0)}_g(t)(1-e^{2\pi\omega_i/\omega_r})$,
and therefore

\begin{equation}
\Phi^{(0)}_g(t+T)=\Phi^{(0)}_g(t)\,e^{2\pi\omega_i/\omega_r}\,.
\label{40}
\end{equation}
Expressions (\ref{39}) and (\ref{40}) are clearly connected to the variation
in time of the axial energy perturbation displayed in Figure (\ref{fig2}).

\section{Polar perturbations of the Schwarzschild black hole}

In this section, we repeat the analysis developed in the previous section
for the axial perturbations. In the context of polar perturbations, the 
calculations are more intricate, in spite of the similarities with the
calculations carried out for the axial perturbations. We will consider the
perturbed metric tensor as given by $g_{\mu\nu}=\bar{g}_{\mu\nu}+h_{\mu\nu}$.
The line element for the unperturbed metric $\bar{g}_{\mu\nu}$ tensor is

\begin{equation}
ds^2=-fdt+f^{-1}dr^2+r^2(d\theta^2+\sin^2\theta d\phi^2)\,,\label{p1}
\end{equation}
and the polar perturbations $h_{\mu\nu}$ are described by \cite{RW}

\begin{equation}
h_{\mu\nu}= e^{-i\omega t}P_l(\cos\theta) \left(
                                              \begin{array}{cccc}
                                                H_0 \,f & H_1 & 0 & 0 \\
                                                H_1 & H_2\,f^{-1} & 0 & 0 \\
                                                0 & 0 & r^2 K & 0 \\
                                                0 & 0 & 0 & r^2K\sin^2\theta \\
                                              \end{array}
                                            \right)\label{p2}\,,
\end{equation}
where $f=1-2m/r$ and 
$H_0$, $H_1$, $H_2$, $K$ are functions of $t,r$. Therefore, the perturbed 
metric tensor is written as

\begin{equation}
g_{\mu\nu}=\left(
             \begin{array}{cccc}
               -f(1-H_0') & H_1' & 0 & 0 \\
               H_1\,' & f^{-1}(1+H_2') & 0 & 0 \\
               0 & 0 & r^2(1+K') & 0 \\
               0 & 0 & 0 & r^2(1+K')\sin^2\theta \\
             \end{array}
           \right)\label{p3}\,,
\end{equation}
where we define
$$H_0'=H_0e^{-i\omega t}P_l(\cos\theta)\,,$$

$$H_1'=H_1e^{-i\omega t}P_l(\cos\theta)\,,$$

$$H_2'=H_2e^{-i\omega t}P_l(\cos\theta)\,,$$

$$K'=Ke^{-i\omega t}P_l(\cos\theta)\,.$$

Einstein's equations reduce to seven non-trivial equations: one algebraic
equation, three first order differential equations for the metric tensor 
components, and three second order equations. The algebraic equation is
eventually reduced to \cite{RW}

\begin{equation}
H_{2}=H_{0}\equiv H\label{p4}\,.
\end{equation}
The three first order differential equations are \cite{Zerilli}

\begin{equation}
\frac{dK}{dr}+\frac{r-3m}{r(r-2m)}K-\frac{1}{r}H+
\frac{1}{2}\frac{l(l+1)}{i\omega r^{2}}H=0\label{p5}\,,
\end{equation}
\begin{equation}
\frac{dH}{dr}+\frac{r-3m}{r(r-2m)}K-\frac{r-4m}{r(r-2m)}H+
\left[\frac{i \omega r}{r-2m}+
\frac{1}{2}\frac{l(l+1)}{i\omega r^{2}}\right]H_{1}=0\label{p6}\,,
\end{equation}
\begin{equation}
\frac{dH_1}{dr}+\frac{i\omega r}{r-2m}K+
\frac{i\omega r}{r-2m}H+\frac{2m}{r(r-2m)}H_{1}=0\label{p7}\,.
\end{equation}
The three remaining equations eventually yield the algebraic equation

\begin{eqnarray}
-\left[\frac{6m}{r}+2\lambda\right]H+\left[2\lambda-
\frac{2\omega^{2}r^{3}}{r-2m}+
\frac{2m(r-3m)}{r(r-2m)}\right]K &{}& \nonumber \\
+\left[2i\omega r+M\frac{l(l+1)}{i\omega r^{2}}\right]
H_{1}&=&0\label{p8}\,,
\end{eqnarray}
where $\lambda\equiv \frac{1}{2}(l-1)(l+2)$.

By rewriting the field quantities according to \cite{Zerilli}

\begin{equation}\label{p9}
K=\alpha(r)\psi+\beta(r)\Theta\,,
\end{equation}
and
\begin{equation}\label{p10}
R\equiv\frac{1}{\omega}H_{1}=\gamma(r)\psi+\rho(r)\Theta\,,
\end{equation}
and defining the tortoise coordinate $x$ as

\begin{equation}\label{p11}
dx=\frac{1}{f}dr\Rightarrow x=r+2m\ln\left(\frac{r}{2m}-1\right)\,,
\end{equation}
we obtain the relations

\begin{equation}\label{p12}
\alpha(r)=\frac{\lambda(\lambda+1)r^{2}+
3\lambda mr+6m^{2}}{r^{2}(\lambda r+3m)}\,,
\end{equation}
\begin{equation}\label{p13}
\beta(r)=1\,,
\end{equation}
\begin{equation}\label{p14}
\gamma(r)=i\frac{-\lambda r^{2}+3\lambda mr+3m^{2}}{(r-2m)(\lambda r+3m)}\,,
\end{equation}
\begin{equation}\label{p15}
\rho(r)=-i\frac{r^{2}}{r-2m}\,.
\end{equation}
It follows that 
\begin{equation}\label{p16}
\Theta=\frac{d\psi}{dx}
\end{equation}
and
\begin{equation}\label{p17}
\frac{d^{2}\psi}{dx^{2}}+\left(\omega^{2}-V_{Ze}\right)\psi=0\,,
\end{equation}
where

$$V_{Ze}= f(r)\frac{2\lambda^2(\lambda+1)r^3+
6\lambda^2mr^2+18\lambda m^2r+18m^3}{r^3(\lambda r+3m)^2}$$
is the Zerilli potential. 

Although the algebraic forms of the Zerilli and
Regge-Wheeler potentials are quite distinct, they share strong similarities
\cite{Nollert}. For very large values of the coordinate $r$, we have 
$V_{Ze}\rightarrow 0$. Thus, in the limit $r>> m$
we obtain a simple solution for the function $\psi$, given by

\begin{equation}\label{p18}
\psi(x)\propto e^{i\omega x} \Rightarrow \psi(x)=Ae^{i\omega x}\,,
\end{equation}
where $A$ is a constant with dimension of length. In order to ensure the 
perturbative character of the solution, the amplitude $A$ is required to
satisfy $A<<1$, in natural units. In the limit $r>>m$, the solution $\psi$
grows with the radial distance. Thus, since the approximations are made 
assuming $|\psi^{2}|<<|\psi|$, we must ensure that 
$|A^{2}e^{2i\omega x}|<<|Ae^{i\omega x}|$. The frequencies of the polar 
perturbations are the same as for the axial perturbations 
\cite{chandrasekhar1975quasi}, provided we consider the same values of
$(n,l)$ when comparing the frequencies. Considering as usual 
$\omega=\omega_{r}+i\omega_{i}$, we have

\begin{equation}\label{p19}
|Ae^{i\omega_{r}x}e^{-\omega_{i}x}|<<1\Rightarrow 
A<<e^{\omega_{i}x}=e^{-|\omega_{i}|x}\,.
\end{equation}
It is important to remark that the stability of the 
solutions demand that the function
$\Psi=e^{-i\omega t}\psi$ decay sufficiently fast with time. Consequently, the
stable solutions are those for which $\omega_{i}<0$. The solutions
obtained in the following approximations are valid as long as
\begin{equation}\label{p20}
x<<\frac{1}{|\omega_{i}|\ln{A}}\,.
\end{equation}
Therefore, the approximate analytic solutions will be valid provided the
condition $-\infty<<x<<\frac{1}{|\omega_{i}|\ln{A}}$ is satisfied.

Just like in the axial case, we must find the asymptotic limits for the
perturbative functions. Thus, in
the limit $r>> m$, equations (\ref{p12},\ref{p13},\ref{p14},\ref{p15})
become 

\begin{equation}\label{p21}
\alpha(r)\approx \frac{\lambda+1}{r}\,,
\end{equation}
\begin{equation}\label{p22}
\beta(r)=1\,,
\end{equation}
\begin{equation}\label{p23}
\gamma(r)\approx -i\,,
\end{equation}
\begin{equation}\label{p24}
\rho(r)\approx-ir\,.
\end{equation}
As a consequence, equations (\ref{p9},\ref{p10}) are approximated by
\begin{equation}\label{p25}
K\approx \Theta = \frac{d\psi}{dx}=
\left(\frac{\lambda+1}{r}+i\omega\right)\psi
\end{equation}
and 
\begin{equation}\label{p26}
H_{1}\approx \omega^{2}r\psi\,.
\end{equation}
Making use of eq. (\ref{p8}) in the limit $r>>m$, together with the
equations (\ref{p25},\ref{p26}), we obtain
\begin{equation}\label{p27}
H\approx -\omega^{2} r\psi\,.
\end{equation}
The expressions above given by eqs. (\ref{p25},\ref{p26},\ref{p27}) will be 
used in the following subsection, where we will evaluate the gravitational
energy of the polar perturbations.

\subsection{The gravitational energy of the polar perturbations}

In similarity to the analysis in subsection 3.1, where we addressed the 
gravitational energy of the axial perturbations, here we also construct
a set of tetrad fields adapted to static observers (with respect to the 
asymptotic flat space-time limit of the black hole), that yields the 
metric tensor (\ref{p3}). A set of tetrad fields that satisfy the necessary
conditions is given by 

\begin{equation}\label{p28}
e_{a\mu}=\left(%
\begin{array}{cccc}
  -A & -B & 0 & 0 \\
  0 & C\sin\theta\cos\phi & Dr\cos\theta\cos\phi & -Dr\sin\theta\sin\phi \\
  0 & C\sin\theta\sin\phi & Dr\cos\theta\sin\phi & Dr\sin\theta\cos\phi \\
  0 & C\cos\theta & -Dr\sin\theta & 0 \\
\end{array}
\right) \,,
\end{equation}
where
\begin{eqnarray}
A&=&f^{1/2}(1-H')^{1/2}\,,\nonumber\\
AB&=&-H_1'\,,\nonumber\\
AC&=&[1-H^{'2}+H_1'\,^{2}]^{1/2}\,,\nonumber\\
D&=&(1+K')^{1/2}\label{p29}\,,
\end{eqnarray}
and whose determinant is $e=D^2ACr^2\sin\theta$. 

The components of the torsion tensor $T_{\lambda\mu\nu}$ and of the tensor
$\Sigma^{\lambda\mu\nu}$ that are needed in the calculations below are

\begin{eqnarray}
T_{212}&=&\frac{1}{2}\partial_1(D^2r^2)-DCr \,, \nonumber \\
T_{313}&=&\frac{1}{2}\partial_1(D^2r^2\sin^2\theta)-DCr\sin^2\theta 
\,, \nonumber \\
T_{202}&=&\frac{1}{2}\partial_0(D^2r^2)\,, \nonumber \\
T_{303}&=&\frac{1}{2}\partial_0(D^2r^2\sin^2\theta)\label{p30}\,,
\end{eqnarray}
and
\begin{eqnarray}
\Sigma^{001}&=&-\frac{1}{2}(g^{01}g^{01}-g^{00}g^{11})
(g^{22}T_{212}+g^{33}T_{313})\,,\nonumber \\
\Sigma^{101}&=&\frac{1}{2}(g^{01}g^{01}-g^{00}g^{11})
(g^{22}T_{202}+g^{33}T_{303})\label{p31}\,.
\end{eqnarray}
With the help of these quantities, we find

\begin{equation}\label{p32}
\Pi^{(0)1}=4k\sin\theta\left[\frac{r^2}{2C}\partial_1(D^2)
+\frac{D^2r}{C}-Dr-\frac{Br^2}{2AC}\partial_0(D^2)\right]\,.
\end{equation}

The expression above leads to the gravitational energy given by eq. (\ref{11}).
An approximate analytic expression for the quantity above may be obtained by
considering terms up to the second order in $H_0$, $H_1$ and $K$. The
unperturbed expression of the momentum given by eq. (\ref{p32}) is simply
$\Pi^{(0)1}_g=4k\sin\theta r(f^{1/2}-1)$. The integration of this quantity 
on a surface at spacelike infinity yields the expected expression for the 
total gravitational energy, $E=m$. Thus, the quantity that yields 
the gravitational energy of the perturbations is 
$\delta\Pi^{(0)1}=\Pi^{(0)1}-\Pi^{(0)1}_g$. The approximate analytic expression
of the latter is

\begin{eqnarray}
\delta\Pi^{(0)1}&\approx &
4k\sin{\theta}\left[\frac{r^{2}}{2}f^{1/2}\partial_{1}K'
+rK'\left(f^{1/2}-\frac{1}{2}\right)-\frac{1}{2}rf^{1/2}H'\right]\nonumber\\
&+&4k\sin{\theta}
\biggl[\frac{r}{2}f^{1/2}\left(\frac{3}{4}H^{'1}-H_{1}^{'2}
-K'H'\right)+\frac{r}{8}K'^{2} \nonumber\\
&+&\frac{r^{2}}{2}f^{-1/2}H'_{1}\partial_{0}K'-
\frac{r^{2}}{4}f^{1/2}H'\partial_{1}K'{}
\biggr]\,.
\label{p33}
\end{eqnarray}
The first line in (\ref{p33}) contains the first order terms 
$\delta\Pi^{(0)1}_{first}$, whereas the other terms correspond to the second 
order terms $\delta\Pi^{(0)1}_{sec}$. The total energy within a spherical
surface of radius $r$ is obtained by
summing these two quantities, $\delta\Pi^{(0)1}=\delta\Pi^{(0)1}_{first}+
\delta\Pi^{(0)1}_{second}$ and integrating over
the spherical surface. The first order
terms depend on the variable $\theta$ in the form $P_{l}(\cos{\theta})$. The
integration of this quantity vanishes, i.e.,
$\int_{0}^{\pi}{d(cos\theta)P_{l}(\cos{\theta})}=0$. Therefore, only the second
order terms yield a non-trivial result when integrated over the whole spherical
shell. 

Using the results given by eqs. (\ref{p25},\ref{p26},\ref{p27}), we integrate 
(\ref{p33}) over a spherical surface of radius $r$, and obtain the 
gravitational energy of the perturbation. We find

\begin{equation}\label{p34}
\delta P^{(0)}\approx -\frac{1}{16}r^{3}\omega^{4}
\Psi(t,r)^{2}\int_{0}^{\pi} P_{l}(\cos\theta)^{2}\sin{\theta}d\theta\,,
\end{equation}
where $\Psi(t,r)=e^{-i\omega t}\psi(r)$.
In view of the orthogonality property of the Legendre Polinomials,

$$\int_{0}^{\pi} P_{l}(\cos\theta)^{2}\sin{\theta}d\theta=\frac{2}{2l+1}\,,$$
we obtain

\begin{equation}\label{p35}
\delta P^{(0)}\approx -\frac{1}{8}\frac{r^{3}\omega^{4}}{2l+1}\Psi(t,r)^{2}\,.
\end{equation}

In similarity to the axial case, the final result given by eq. (\ref{p35}) has
real and imaginary parts, both in the exponential and in the frequency $\omega$.
As before, we take the real part of the whole expression. In the case of 
highly damped oscillations, we have $\omega_{r}<<\omega_{i}$, and therefore
we may use $\omega \approx \omega_{i}$.

\begin{figure}[htbp]
\centering
	\centering
		\includegraphics[width=1\textwidth]{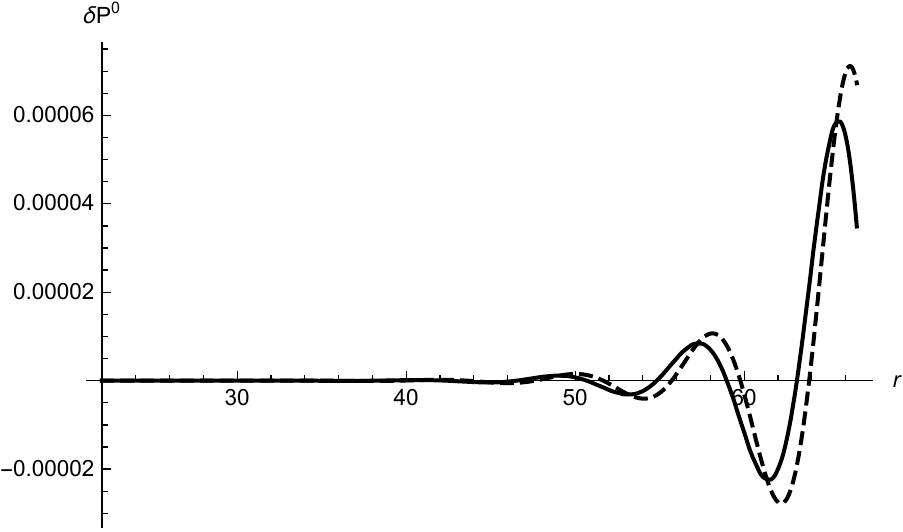}
	\caption{Comparison between the real parts of the gravitational energy of the 
	polar perturbation. The continuous line represents the numerical value of the 
	evaluation, and the dashed line corresponds to the value resulting of the 
	approximate analytic expression (\ref{p35}). 
	The parameters used in the analysis are
	$A=10^{-6}$, $m=1$, and the frequency considered is the fundamental one
	($n=0,l=2$) $\omega=0.373162-i0.089217$, obtained by means of the third order
	WKB method. The data represent the initial instant of time of the
	perturbation, i.e., $t=0$.}\label{fig3}
\end{figure}

\begin{figure}[htbp]
	\centering
		\includegraphics[width=1\textwidth]{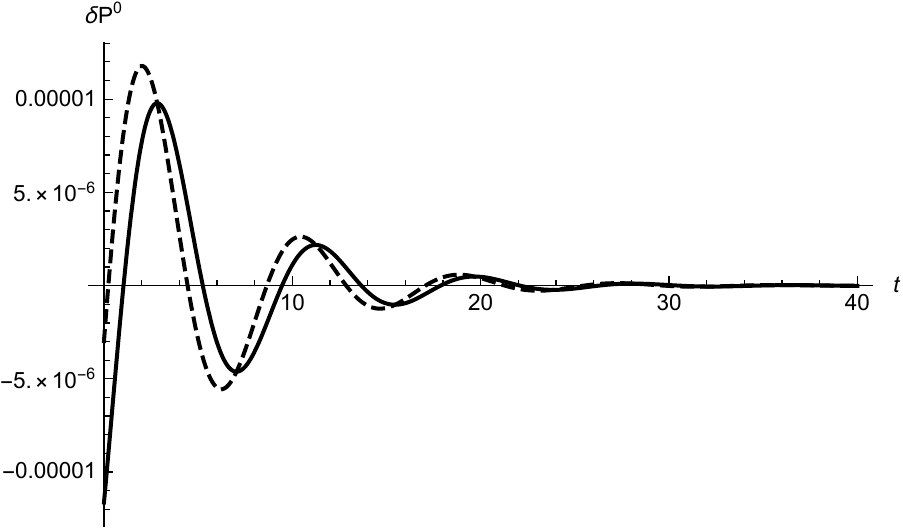}
	\caption{Comparison between the real parts of the gravitational energy of the 
	polar perturbation. The continuous line represents the numerical value of the 
	evaluation, and the dashed line corresponds to the value resulting of the 
	approximate analytic expression (\ref{p35}). 
	The parameters used in the analysis are
	$A=10^{-6}$, $m=1$, and the frequency considered is the fundamental one
	($n=0,l=2$) $\omega=0.373162-i0.089217$, obtained by means of the third order
	WKB method. The integration is made on a surface of radius $r=60$, in natural
	units.}\label{fig4}
\end{figure}

The solution given by eq. (\ref{p35}) may be compared with the one obtained
directly from the numerical integration of eq. (\ref{p32}). In Figure 
\ref{fig3}, the continuous line represents the numerical solution, whereas 
eq. (\ref{p35}) is represented by the dashed line. In the numerical solution,
the function $\psi$ was obtained by considering the full expression of the 
Zerilli potential, and the boundary conditions are exactly the same as 
considered in the case of axial perturbations. As before, the numerical 
solution was
obtained considering $r_{1}=2m+m/4$ and $r_{2}=200m$ in the boundary conditions
for eq. (\ref{p17}). We believe that the errors in the approximate analytic
solution, compared to the numerical solution, are due to the approximation
made in eq. (\ref{p17}), where we neglected the Zerilli potential in the 
limit $r>> m$. In Figure \ref{fig3}, we are considering the initial instant
of time of the perturbation. The time evolution of the energy of the
perturbation is stable, since the gravitational energy of the perturbation
is damped in the course of time, as wee see in Figure \ref{fig4}.


\section{Final remarks}

In this paper we obtained the expression of the gravitational energy 
of the perturbations due the quasinormal modes in the Schwarzschild space-time.
The expressions given by (\ref{34},\ref{p35}) hold in the limit 
$m<< r << \infty$, and depend on the integers $n$ and $l$. The variation
in space and in time of these quantities are displayed in Figures
(\ref{fig1}-\ref{fig4}). These are the major results of the present 
analysis. The dependence on the coordinate $r$ in expressions 
(\ref{34},\ref{p35}) is due to the very nature of the QNM. The functions $h_0$
and $h_1$ diverge at spacelike infinity, which is a feature of the QNM. 
However, the perturbation is assumed to be localized, and 
does not make sense in the limit $r\rightarrow \infty$. In our understanding,
the axial and polar quasi-normal perturbations represent ripples in the
space-time geometry, as they do represent in the metric formulation of 
general relativity. However, the physical description and results displayed
by Figures (\ref{fig1},\ref{fig2},\ref{fig3},\ref{fig4}) can only be obtained
in  the TEGR, since the the gravitational energy-momentum $P^a$ given by 
eqs. (\ref{8}) and (\ref{11}) cannot be established in the standard 
formulation of general relativity.

One conclusion is that the gravitational energy of the black hole oscillates.
However, since the energy of the perturbations is concentrated far from the
event horizon, as we see in Figures (\ref{fig1},\ref{fig3}), we conclude
that the {\it mass} of the black hole does not vary in time, i.e., the mass
of the black hole (restricted to interior of the event horizon)
neither increases nor decreases with time. Figures 
(\ref{fig2},\ref{fig4}) show that energy of the perturbations rapidly 
decays for large instants of time.  Here we make a distinction between the 
gravitational energy of the black hole (the zeroth component of the 
gravitational
energy-momentum 4-vector $P^a$), and the mass of the black hole (an invariant 
of the Poincar\'e group, $P^aP_a = -m^2$). Note, in addition, that $m$ is the
mass of the black hole established in a frame where the black hole is at rest.
In view of these considerations,
we may conclude that the energy of the perturbations is a non-local effect,
since it takes place sufficiently far from the event horizon.
We may conjecture that the farther one is from the black hole, i.e., the
farther an external perturbation is imparted to the black hole, the more energy
is required to perturb it.

The dependence of expressions (\ref{34},\ref{p35}) on the integers $n$ 
and $l$ implies a discretization of the energy perturbations. This 
discretization must be further investigated. However, it is known that the 
very discretization of the QNM leads to a 
discretization of the area of the black hole's event horizon \cite{Maggiore}.
A possible interplay between the discretizations of the energy perturbations
and of the area of the black hole is an issue of relevant interest.

The flux of gravitational radiation emitted in the process of damping
of the quasinormal modes of a Schwarzschild black hole has been 
investigated in refs. \cite{CPM,Nagar,Nakano}, by means of the
Landau-Lifshitz pseudotensor \cite{LL}, assuming that at large
distances from the black hole the space-time perturbation is 
represented by a plane gravitational wave. The analysis in the latter
references was made assuming $n=0\;,l=2$. 
In the present investigation, the 
expression for the flux of gravitational radiation given by eq. (\ref{9})
is invariant under coordinate transformations, in contrast to the
pseudotensors, that are coordinate dependent expressions. It would be
interesting to address a specific physical configuration like the collision
of two black holes, and compare the energy radiated after the collision (i) 
by means of eq. (\ref{39}), and (ii) through the approach based on the latter
references, in the context of the Landau-Lifshitz pseudotensor. This issue
will be addressed elsewhere.

\begin{figure}[htbp]
	\centering
		\includegraphics[width=.8\textwidth]{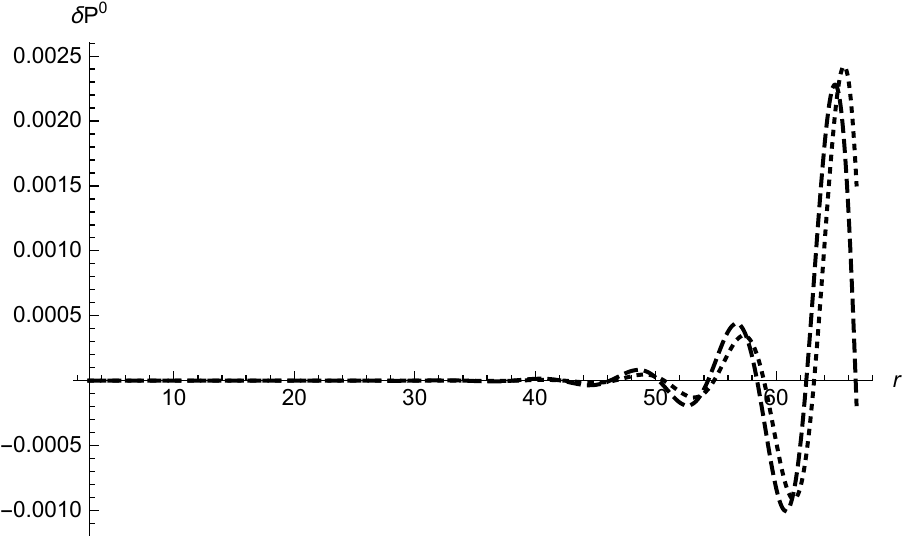}
	\caption{Comparison between the real parts of the gravitational energy of the
	perturbations. The dashed line represents the numerical expression of the 
	energy for the axial perturbations, and the dotted line represents the 
	numerical energy
	of the polar perturbations. In natural units (whenever necessary), we have
	$A_{axial}=10^{-6}$, $A_{polar}=5^{-1/2}10^{-6}$, $m=1$, and the frequency 
	(fundamental, ($n=0,l=2$)) is $\omega=0.373162-i0.089217$, obtained by means
	of the third order WKB method. The data represent the initial instant of
	time of the perturbations, i.e., $t=0$.}\label{fig5}
\end{figure}

The amplitude $A$ in the function $\psi$ arises as a constant of integration
in the solutions of eqs. (\ref{19}) and (\ref{p17}). Let us consider the
tortoise coordinate $x=r+2M\ln{\left(\frac{r}{2M}-1\right)}$. If we make
$x'=x+\frac{1}{i\omega}\ln{A}$, then it is easy to see that 
$\psi=e^{i\omega x'}=Ae^{i\omega x}$. Therefore, there is an arbitrariness 
in the establishment of the amplitude $A$. Since the profiles of the 
gravitational energy perturbations in Figures (\ref{fig1}-\ref{fig4}) is
very much similar, with a suitable choice of amplitudes we may locally 
enhance the similarity between the two perturbations, as we see in Figure 
(\ref{fig5}). However, note that $\psi$ is dimensionless in the axial
perturbation, but has dimension (of distance) in the polar perturbation.
But still, the similarities between the two perturbations is very clear.

The procedure presented in this article may also be applied to extended 
formulations of gravity, namely, the f(T) type theories. In some of these 
approaches there are proposals for the gravitational energy, as
for instance in refs. \cite{Capozziello-1}, \cite{Capozziello-2}, 
\cite{Capozziello-3}, \cite{Capozziello-4}, \cite{SC-Ulhoa}.
There are various motivations to address such
extended theories, as for instance the analysis of the dark energy and dark
matter problems, the problem of singularities of black holes, and the 
investigation of the degrees of freedom carried by gravitational waves. The
analysis of the gravitational energy perturbations in these extended 
frameworks is, of course, of relevant interest.

\end{document}